\documentclass[numbers,10pt]{sigplanconf}

\usepackage{oopsla2016}
\usepackage{graphicx}
\usepackage{url}
\usepackage[utf8]{inputenc}
\usepackage[T1]{fontenc} 
\usepackage[table,dvipsnames]{xcolor}
\usepackage{listings}
\usepackage{microtype} 
\usepackage[bookmarks=false,hidelinks]{hyperref}
\usepackage{booktabs}
\usepackage{flushend}

\lstdefinelanguage{SpecElektra}{
        basicstyle=\ttfamily\small,
        comment=[l]{;},
        commentstyle=\color{purple}\ttfamily,
        morestring=[b]',
        morestring=[b]`,
        morestring=[b]",
        stringstyle=\color{red}\ttfamily,
        sensitive=f,
        keywordstyle=\color{BlueViolet}\bfseries,
        keywordstyle=[2]\color{Melon}\bfseries,
        keywordstyle=[3]\color{Aquamarine}\bfseries\textit,
        keywordstyle=[4]\color{NavyBlue}\bfseries,
        keywordstyle=[5]\color{Mahogany},
        keywords={context, property1, property2, file, content},
        keywords=[2]{layer},
        keywords=[3]{name, order, interface, network, emphasized},
        keywords=[4]{[, ]},
        keywords=[5]{getenv, open},
        literate={=}{{{\color{red}\textbf=}}}1
                 {<-}{{{\color{red}\textbf<-}}}2
                 {:=}{{{\color{red}\textbf:=}}}2
                 {*}{{{\color{green}\textbf*}}}1
                 {\%}{{{\color{NavyBlue}\textbf\%}}}1
                 {[}{{{\color{Sepia}\textbf[}}}1
                 {]}{{{\color{Sepia}\textbf]}}}1,
}

\makeatletter
\def\addToLiterate#1{\edef\lst@literate{\unexpanded\expandafter{\lst@literate}\unexpanded{#1}}}
\lst@Key{add to literate}{}{\addToLiterate{#1}}
\makeatother

\newcommand{\cfgfile}[1]{
\lstinputlisting[basicstyle=\ttfamily\small,linerange=#1-end]{code.cpp}
}

\newcommand{\srcfile}[1]{
\lstinputlisting[basicstyle=\ttfamily\small,linerange=#1-end,language=C++,literate=]{code.cpp}
}

\lstMakeShortInline|

\makeatletter
\newcommand{\property}[1]{\@property{#1}}
\newcommand{\@@property}[1]{\textsc{\let\@property\@@@property#1}}
\newcommand{\@@@property}[1]{\textnormal{\let\@property\@@property#1}}
\let\@property\@@property
\makeatother

\makeatletter
\newcommand{\keyname}[1]{\@keyname{#1}}
\newcommand{\@@keyname}[1]{\texttt{\let\@keyname\@@@keyname#1}}
\newcommand{\@@@keyname}[1]{\textnormal{\let\@keyname\@@keyname#1}}
\let\@keyname\@@keyname
\makeatother

\makeatletter
\newcommand{\strong}[1]{\@strong{#1}}
\newcommand{\@@strong}[1]{\textbf{\let\@strong\@@@strong#1}}
\newcommand{\@@@strong}[1]{\textnormal{\let\@strong\@@strong#1}}
\let\@strong\@@strong
\makeatother

\title{Persistent Contextual Values as Inter-process Layers}

\authorinfo{Markus Raab}
{Institute of Computer Languages \\
Vienna University of Technology, Austria
}
{markus.raab@complang.tuwien.ac.at}

\begin{document}

\toappear{}
\toappear{

\noindent
{\scriptsize
Mobile!’16, October 31, 2016, Amsterdam, Netherlands 
}
}

\maketitle

\begin{abstract}
Mobile applications today often fail to be context aware when they also need to be customizable and efficient at run-time.
Context-oriented programming allows programmers to develop applications that are more context aware. 
Its central construct, the so-called layer, however, is not customizable.
We propose to use novel persistent contextual values for mobile development. 
Persistent contextual values automatically adapt their value to the context.
Furthermore they provide access without overhead.
Key-value configuration files contain the specification of contextual values and the persisted contextual values themselves.
By modifying the configuration files, the contextual values can easily be customized for every context.
From the specification, we generate code to simplify development.
Our implementation, called Elektra, permits development in several languages including C++ and Java.
In a benchmark we compare layer activations between threads and between applications.
In a case study involving a web-server on a mobile embedded device the performance overhead is minimal, even with many context switches.
\end{abstract}



\section{Introduction}
\label{introduction}

Context-oriented programming (COP) aims at software modularization, with focus on considering contextual behavior~\cite{hong2009context}.
It enables us to implement context-aware behavior separately.
Central to COP are \emph{layers}, i.e.\ the modules for such code.
Active layers encode the current context of an application.

\emph{Contextual values} (CVs) are variables whose values depend on context, i.e.\ active layers.
CVs fit nicely into COP ideas because of their seamless integration with layers.
Side-effects of CVs are limited to their respective context.
Their main advantage is their simplicity because CVs ``boil down to a trivial
generalization of the idea of thread-local values''~\cite{tanter2008contextvalues}.

Currently many COP languages do not provide straight-forward, context-aware activation of layers.
Instead they mainly support code-snippets for activation~\cite{kamina2015generalized}, e.g.:
\begin{lstlisting}[language=C++,literate=,numbers=none,xleftmargin=0.0ex]
activate (Austria) if (gps_pos=="austria")
\end{lstlisting}
In this example, we activate the layer |Austria| according to the current GPS position.
Developers can easily miss some context information at some places, considering that activations are spread across the whole source code.

Another problem COP languages currently face is that layer information cannot easily be shared between applications.
Solutions with context-oriented middleware are often not light-weight and need special tools for introspection and debugging.
We propose to persist layer information into configuration files.
While it is well-known how to persist objects, we lack similar techniques for layers.
Because of this issue, it is difficult to separate context sensors and applications.
Often the same sensed context, however, is needed in several applications which leads to duplication of efforts.

In this paper we describe the novel idea to use persistent contextual values for layer activation.
Our goal is to combine layer activations with CVs that are shared across applications.
In our approach we will use CVs as parameters for |activate| and |with| constructs.
COP languages commonly use these constructs for layer activation~\cite{kamina2015generalized}.

Our main contribution is a fully implemented framework fulfilling the above goals. 
It has already successfully been used in embedded systems.
It does not make any assumptions of the application's architecture.
Our implementation provides 
(1) generated CVs to be used like variables, and
(2) initialization, persistency, and notification for them.
It supports many languages, including C++ and Java.   

Because CVs are context-aware they will automatically consider their context.
By persisting CVs we synchronize layer activations between applications.
Our technique even works across different programming languages.

To be better suitable for mobile development one additional goal is that accessing CVs should not add overhead.
Instead it should deliver the same performance as reading native variables.
Thus our approach relies on explicit activation and a cache for every CV~\cite{raab2014program}.

We avoid the |if| in the |Austria| layer activation above.
Instead we would activate the contextual value |position|:
\begin{lstlisting}[language=C++,literate=,numbers=none]
activate (position)
\end{lstlisting}
This way, due to CV semantics, contexts of |position| are automatically considered.
The contextual value would be persistent and shared between applications.
When a context sensor has new information available, other applications would be notified via our framework.

For example, think about internationalized software.
CVs help us to easily show correct translated messages.
Currently, however, we have to activate the correct layers in every COP application individually.
There was no easy way for one application to tell all other applications that the language has changed.
We want to be able to activate layers with CVs:

\srcfile{visitShort}

A code generator yields the classes |Person| and |Language|.
They implement CVs semantics.
In the Line 2 we activate the contextual value |lang|.
A previously persisted value initializes CVs at application startup.
This initialization provides sensible defaults changeable by user settings or context.

We answer the following research questions:
\begin{description}
\item[RQ1]: How can contextual values be used for layer activation? Which limitations does it have?
\item[RQ2]: What are the costs of inter-process layer activation?
\end{description}

The questions are significant because it is common that we face a number of applications and programming languages in our system.
Such a combination, however, currently is not well supported by COP.
Our idea is not only feasible but practical as demonstrated by a prototype.

The paper is structured as follows:
In Section~\ref{background} we explain the background.
In Section~\ref{approach} we elaborate our approach
and in Section~\ref{evaluation} we evaluate it.
After considering related work in Section~\ref{related}, we conclude in Section~\ref{conclusion}.

\section{Background and Syntax}
\label{background}

We implemented our approach as tool called Elektra.
In this section we will explain the essence and syntax previous versions of Elektra.
Earlier versions lacked possibilities to use CVs for activations and needed extra layer specifications. 

CVs are very useful to interact with a program execution environment (PEE), i.e.\ configuration files and environment variables.
CVs ensure that the context always is taken into account when accessing the PEE.
PEE is an elegant way to persist CVs: we write key-value pairs into configuration files.

For PEE the performance focus lies on retrieving values. 
Modification of the PEE is usually only done manually by users when they change settings.
Thus, when changing PEE, proper validation is more important than performance.
In earlier work we demonstrated that PEE can be tightly integrated with CVs~\cite{raab2015global,raab2014program}.
We propose the adoption of PEE as CV storage and maintain the goal to have fast access when reading CVs.

In our approach, a small library abstracts from syntax and location of the configuration files.
Elektra uses the PEE, such as configuration files, to initialize contextual values.
Elektra supports over 190 configuration file formats, including INI, XML and JSON formats.
In this paper we will use a simple key-value syntax to illustrate the content of the key-value database. 
To easily distinguish configuration and its specification in this paper, we use:
(1) assignment with |=| for configurations containing the values of every CV,
(2) keys written in |[]| and assignment with |:=| only for specifications:

\cfgfile{syntax}

In the first line of the example above we \emph{configure} the option identified with the key |path/key|.
As hierarchy separator we use |/|.
The key |key| is below |path| and has the value |value|.
In lines 3 and 4 we specify two properties for the same key |path/key|:
they are called |property|$N$ and have respective values |propvalue|$N$.
The key, value and the properties are stored in the key-value database, for example:

\cfgfile{configfile}

In this example, the CV has 3 different possible values with |*| as wildcard expression.
Furthermore, within the same configuration files, we can also \emph{specify} contextual values.
For example, we specify the CV |greeting|:

\cfgfile{specfile}

The respective key in Line 1, i.e.\ the full string within |[]|, is a contextual value's name.
Lines 2-3 further specify the contextual value with |:=| assignment.
Here we specify that the CV |greeting| has the type |string|.
In Elektra a code-generator synthesizes context-aware classes using contextual value specifications.
The tool generates the code for the underlying CV-classes |Person| and |Greeting|.

Often it is useful to give layers a name~\cite{kamina2015generalized}.
Our approach consistently gives every layer a name, written within |[]|.
All generated classes are nested in one hierarchy with |/| as root.

A single CV has many values for each context.
In the specification strings enclosed in |
For contextual interpretations we substitute the placeholders with values given by layers.
Unique keys to lookup individual values are determined by substituting all placeholders with values from the layers.
We use these keys to lookup values in the configuration files.

When no layer was found, the |*| in the configuration file will match.
The character means that the layer is inactive or empty.
In the above example, if the only active layer is |language| with the value |german|, an instance of the class |Greeting| has the value |Guten Tag!|.

Developers directly utilize the contextual values in their own code.
CVs are used in the same way as variables.
As example we extend our previous C++ snippet:

\srcfile{visit}

The |Person|-object |p| is a contextual value passed via reference parameter on Line 1.
Dynamically scoped context is specified via the |with| construct in lines 2 and 3.
|CountryAustria| and |LanguageGerman| are layers, but not persistent contextual values.
Within the dynamic scope of the block after |with| statements, the content of CVs can differ.

For example, when we modify the nested contextual value |visits| in Line 4, the changes are only visible within the |with| block.
In Line 8, the previous values are restored.

Our approach has introspection capabilities.
We easily can inquiry layer information as done in Line 6.
The introspection is useful for debugging and assertions~\cite{raab2014program}.

The issue with the former approach is the implementation of the layers used in lines 2 and 3.
The previous approach forced us to implement a layer for every contextual variation.
Developers needed to manually implement features such as thread safety, contextual awareness and persistence for every layer.
With many layers for highly-dynamic context-aware applications their implementation can be a burden.
In this paper we describe how we can avoid this extra effort and exclusively use CVs specified in configuration files.

\section{Inter-process Layers}
\label{approach}

We extend Elektra with the possibility to directly activate CVs.
We consider a layer to be active, when a CV is non-empty.
As side-effect, this idea enables inter-process activation of layers because of the persistency of CVs.
To synchronize layers with CVs we lack two essential features:

\begin{itemize}
\item We need an intra-process notification which allows us to update the context of CVs when other CVs, representing layers, change.
\item We need an inter-process notification to know when to reread configuration files.
\end{itemize}

\subsection{Contextual Activation}

One of the main benefits of activation of CVs is that we automatically get contextual activation.
We do not have to worry for every activation if every dependent context is considered.
For example, we have the following CVs:

\cfgfile{locationcv}

And we want to successively activate these CVs:

\srcfile{visitContext}

If |location| and |country| were layers as described in background exactly these layers would be activated in the given order.
This means that the activation in Line 3 will not consider the location activated in Line 4.

But because |location| and |country| are CVs, they take context into account and the activations influence each other.
In this example after the activation of |location|, |country| will be updated according the new position.
Line 4 will also update the |Country| layer according to the established context.

For example, if |location| is an empty string, no layer is activated.
With symbolic links in the CVs specification~\cite{raab2015kps} one can implement even more complex scenarios.
In Line 5 the greeting will be according to local customs or some default if the CVs |country| and |location| are inactive.

The value of CVs usually is determined by context sensors that run as separate active processes.
Their task is to track low-level sensor values such as GPS and pool them to high-level context as needed by other processes.
The main advantage of this approach is the reuse of context information and the decoupling between processes.

\subsection{Specification}

For convenience we decided that by default the last part (separated with |/|) of the key is the layers' name.
We already established that the keys are specified within |[]|.
This convention gives most CVs an appropriate layer name:

\cfgfile{layer}

In above example, the CV-class |Country| will automatically have the layer name |country| when activated.
In some situations the last part of the name is not the right choice.
For example, we use a country code to determine a country:

\cfgfile{layername}

Layer names, unlike CV-classes, do not provide a hierarchy.
We likely do not want the layer to be named |code|, thus we rename it to |country| as specified by |layer/name|.
Then activation of such a CV activates the layer |country|.

\subsection{Intra-process Notification}

In previous versions of Elektra it was assumed that every change of a CV is caused by the assignment to a CV.
In this extension we avoid this assumption.
We introduce a reload mechanism when the underlying persistent CVs change.

We implement such a mechanism by an intra-process notification.
For this functionality we use the observer pattern.
The CVs act as observers, the context is the concrete subject.
For in-memory synchronization the method |sync| can be used:

\srcfile{sync}

In Line 2 we fetch all values for every context from the configuration files.
In Line 3 we call the in-memory update of all CVs to reload their value.
The |sync| invocation also makes sure that the correct layers are active.

The persistency layer has two further methods for synchronization in both directions:
With |kdb.get| modified configuration files are parsed, with |kdb.set| changes are written to the configuration files.
This way the application developer decides about the behavior in the case of conflicts.
Because a three-way merge is the most-requested behavior, a convenience API exists for this case.

Because we want CVs to act as layers we cannot update the observers in an arbitrary order.
Instead we need to consider the dependencies between CVs.
CVs with placeholders depend on CVs that have the placeholders name as their layer name.
For example, |Country| needs to be updated before |Location| because |Country| has |

We start by updating CVs that do not have dependencies.
Then we update CVs that are dependent on layers that were updated before.
Elektra solves this ordering problem with a topological sort based on Kahn~\cite{kahn1962topological}.

Because in our approach hooks can be registered to be executed on layer activation~\cite{raab2015global}, users might need a specific order.
Thus users can describe a specific order of activation.
For example:

\cfgfile{layerorder}

In the example, no dependency is given.
Thus any order would be correctly topological sorted.
But because of the users preference in |layer/order|, |country| is activated before |language|.
This way the user can be sure that |country| hooks would be executed before |language| hooks.
If the |layer/order| conflicts with topological sort, |layer/order| can only be fulfilled partly and a warning is emitted.

With layers and CVs as completely separate concepts cycles were not possible:
CVs depended on layers, but not the other way round.
An issue of our approach is that we introduce potentially cyclic dependencies.
For example:

\cfgfile{cycle}

If we activate |country| we would need the value of the current |language| layer and vice-verse.
Note that |layer/order| introduced before does not help with this issue.
With specific values stored, every activation leads to toggling values:

\cfgfile{configcycle}

Such cycles usually stem from design errors and are unwanted.
In our approach we prohibit such cycles.
We introduce a limitation that causes some previously valid specification files to be rejected.
These cases can already be detected when parsing the specification.
If the specification nevertheless is faulty, the user will receive a run-time exception.

\subsection{Synchronization Points}

Because CVs are frequently accessed, we want to avoid any overhead when reading the value of CVs.
We achieve this behavior by requiring the developer to define synchronization points.
At synchronization points new values are pushed to CVs using the observer pattern.
Only during synchronization points performance overhead occurs.
Otherwise reading CVs has the same overhead as accessing native variables~\cite{raab2014program}.
Another advantage of explicit synchronization points is that the user has full control over costs occurring in the program.

Making synchronization points explicit might seem to be cumbersome to program.
But in practice it is often obvious where synchronization should occur: when a user starts a new interaction.
With a use-case-based software engineering approach one can systematically find all such places.
Forgetting about a synchronization point will only affect the specific interaction.
In mostly single-threaded applications, which do not sense context itself, it is even simpler:
one only synchronizes the main-thread when the application is notified.

The synchronization points define when intra-process updates will take place.
The context will push all changes to the respective CVs.
For example:

\srcfile{syncPoint}

In the example above, we introduce a synchronization point in Line 2.
All mentioned reloading features only happen during this invocation.
The contextual value |a| is not modified by context changes or changes of persistent CVs at other places.
This means, that the programmer can be sure that |a| is not changed during the loop starting on Line 3. 

\subsection{Assignment}

Another property of our approach is that after an initial activation, every (de)activation can occur via changing the values of CVs:

\srcfile{assignment}

The precondition that layers are influenced via assignments is fulfilled in Line 2.
In Line 3 we see an assignment to an empty string.
Layers with an empty value influence CVs in the same way as deactivated layers.
But only after explicit deactivation in Line 7, changes of the CV |lang| do not influence other CVs anymore.

Synchronization via |sync| activate or deactivate layers in the same way as the assignment does.
One can think of |sync| as correctly ordered assignment of every CV.

\subsection{Inter-process Notification}

Because of diverse requirements we took care that Elektra follows very modular design principles~\cite{raab2010thesis,raab2016improving}.
The inter-process notification requirements differ from system to system.
We decided to implement inter-process notification in plugins.

Whenever a process modifies the underlying configuration files plugins take care of notification.
It is trivial to include notification mechanism that already implement a message bus.
In the plugin you only have to publish the message without any further concerns.
In every interested process one has to implement a listener that fetches the updated configuration.
After every thread has passed a synchronization point, the application is fully updated to the new configuration.

Implementing an inter-process notification without a message bus is more challenging.
Nevertheless such an endeavor is useful for legacy applications.
Notification via signals are very popular because they are part of C89 and POSIX.

The idea is that every process using Elektra registers its process identifications (PIDs) at startup.
Whenever configuration files change, a plugin sends a signal to all registered PIDs.
Within individual applications one has to install a signal handler.
The signal handler is limited to atomic changes.
Thus it can only flag that such an event occurred and other threads can pickup the changes later.

The update to current persistent CVs themselves is via |kdb.get| as shown before.
On conflicts Elektra usually uses a three-way merge to not lose data on concurrent updates of other processes.

\section{Evaluation}
\label{evaluation}

We benchmarked Elektra
on a hp\textsuperscript{\textregistered} EliteBook 8570w using the
central processor unit (CPU)
Intel\textsuperscript{\textregistered}
Core\textsuperscript{\texttrademark} i7-3740QM with 4 cores @ 2.70GHz.
The operating system was Debian GNU/Linux Jessy 8.4 amd64.
We used the compiler gcc \mbox{4.9.2-10}.

\subsection{Microbenchmarks}

\begin{figure}[htp]
\centering
\includegraphics[scale=0.52]{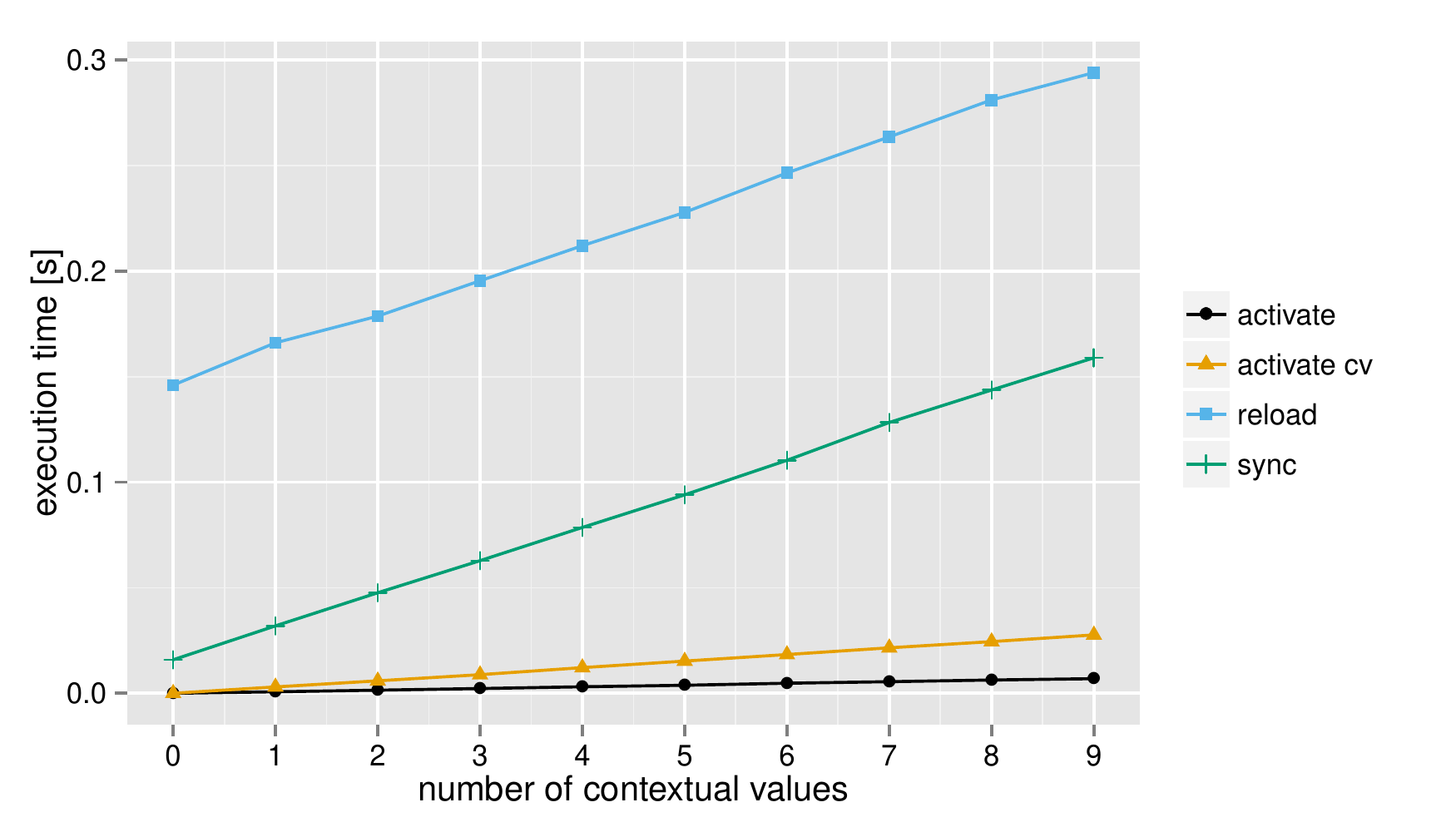}
\caption{Comparison of 1000 iterations in four setups:
by using reloading from persistent storage (reload),
syncing all CVs in memory (sync),
activation of CVs (activate CV),
and only switching layers (activate).
We vary the number of activations or CVs.}
\label{fig:switch}
\end{figure}

We start with microbenchmarks that measure the cost of different synchronization methods.
As we see hard coded in the code snippets below, we used 1000 iterations.
We only show the mean value of the 11 measurements we did for every microbenchmark.
We created four microbenchmarks for each line in Figure~\ref{fig:switch}.
We designed the microbenchmarks in a way that they are valuable for the decision when which activation strategy should be used.
First we start to explain the individual microbenchmarks, then discuss the results.
For all four microbenchmarks we will use the same setup:

\srcfile{benchmarksetup}

The first test, called |benchmarkActivate|, measures activation with layers.
In Line 2 we start the measurement and stop it at Line 10.
In lines 5-7 the relevant action takes place.
In Line 8 another contextual value gets accessed to avoid too aggressive compiler optimizations.
Note that reading Elektra CVs is without any overhead compared to access of native variables, so the Line 8 does not influence the benchmark measurable:

\srcfile{benchmarkactivate}

In the second microbenchmark (|benchmarkActivateCV|), we used the CV-activation feature introduced in this paper.
In the relevant lines 5 to 7, we now have activation of $0$ to $N$ CVs.
In this benchmark, activation of layers only happens implicitly.
Note that again the numbers of activations increase with the number of CVs:

\pagebreak

\srcfile{benchmarkactivatecv}

As third benchmark, we used the |sync| feature introduced in this paper.
The synchronization in Line 6 will synchronize all $N$ different CVs.
In this benchmark we do not reload from persistent storage.
Nevertheless, the implementation must recalculate every activation of every CV.

\srcfile{benchmarksync}

In the last microbenchmark, we fully reload configuration files from disk.
The challenge of this benchmark is an optimization of Elektra.
The operation to fetch data from persistent storage would not reload without changes.
Thus we used for every iteration a different handle to fetch from persistent storage (lines 2 and 3).
This ensures that the configuration file is actually reread:

\srcfile{benchmarkreload}

We see in Figure~\ref{fig:switch} that the overhead increases linearly the more CVs or layers are involved.
More flexible activation scenarios are more expensive.
Especially, reloading from configuration files introduces a relatively large, but constant, offset.
Note that we used a text configuration file parser.

\subsection{Case Study}

In a case study we implemented a web server that outputs localized HTML pages.
Remote users can connect to the web server that is installed on an embedded device.
For localization and session handling the web server heavily relies on CVs.
Because of the features CVs provide, these parts were trivial to implement.
Additionally the web server works together with context sensors that modify persistent CVs.
For example, one sensor detected motion.

During implementation we found CVs very useful.
The specification file was 87 lines and contained 17 CVs.
From this specification file 3623 lines of code defining all CVs as classes in one large hierarchy were generated.
The boilerplate code (lines 1 to 10) left to write is minimal:

\srcfile{main}

In Line 12 we see how we can access CVs in the hierarchy.
We call the function we defined earlier in the paper.

For the benchmark we created one additional thread.
In the thread the web server periodically syncs with persistent changes in the same way as |benchmarkReload| does.
We found out that beginning with 2200 requests/seconds the reply/seconds did not increase significantly anymore.
To measure if context changes influence the functionality of the web server we used httperf on the same machine |l|:

\pagebreak

\srcfile{httperf}

\begin{figure}[htp]
\centering
\includegraphics[scale=0.52]{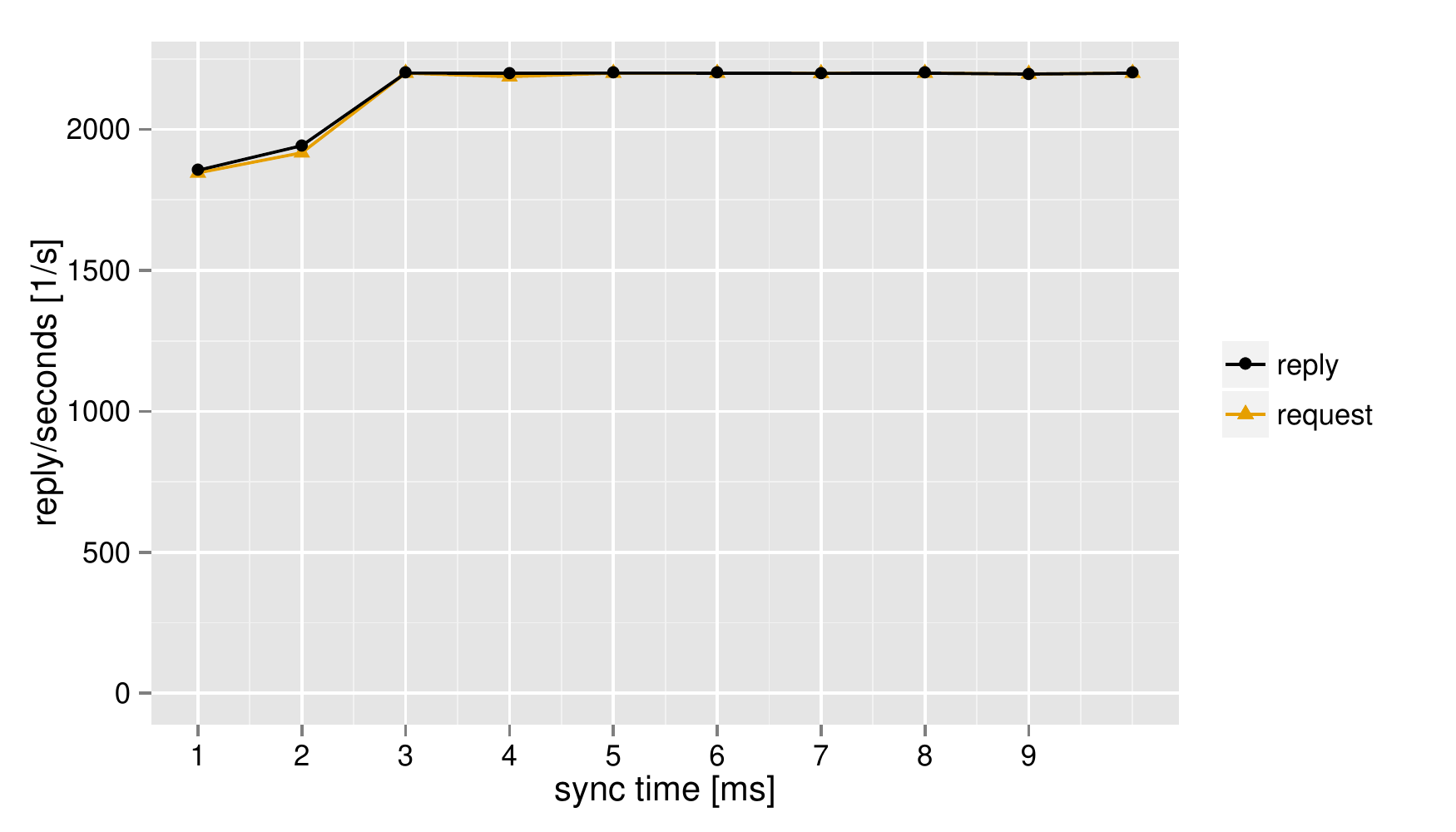}
\caption{Request/reply rate of a web server. The sync time is a sleep interval in milliseconds to wait for the next synchronization with persistent CVs.}
\label{fig:lock}
\end{figure}

As you see in Figure~\ref{fig:lock} the sync rate barely has influence on the request and replies the web server can deliver.
Only at sync rates below 3 ms there is an effect.
A possible explanation is that other CPU cores can compensate easily as long as the required lock during |sync| does not dominate.
Timeouts of requests were nearly always 0 except for sync rates of about 3 ms or when influencing the setup.
We see that Elektra can be practically used for embedded applications and that the number of context changes only has minimal effect, even across applications.

\section{Related Work}
\label{related}

Löwis et al.~\cite{lowis2007contextbeyond} also implemented CVs.
They call them context variables.
Their layer activation (they call it binding of CVs) requires the programmer to explicitly declare layers.
Thus their approach is very similar to previous versions of Elektra and would benefit from the approach described here.

Kamina et al.~\cite{kamina2015generalized} proposed a generalized activation mechanism based on contexts and subscribers.
Their implicit activation, however, has severe impact on the performance. 

Very early work (1979) on contextual values is done by Asirelli et al.~\cite{asirelli1979flexible}.
Their contextual values are depending on state.
We could not find specific information, but 
their system may have achieved something similar to Elektra.
Different from Elektra, they use context-value pairs which do not encode variability information in their keys.
Montangero et al.~\cite{montangero1975magma} described garbage collection techniques for CVs.

Wang et al.~\cite{wang2009context} proposed a new metric useful for testing context-aware applications.
Information present in the specification of Elektra helps to improve testing, too.

Elmongui et al.~\cite{elmongui2009chameleon} described context-aware DBMS.
Because Elektra focuses on persisting CVs the approa\-ches seem to be complementary:
Such query languages could be an extension for some use cases of Elektra.

We also proposed to move the context awareness to a key-value database~\cite{raab2016unanticipated}.
Using interception techniques, unmodified applications were made more context aware.

Mens et al.~\cite{mens2016taxonomy} created a taxonomy of context-aware software variability approaches.
They explicitly mention the execution environment to be an important source for context.
Based on the taxonomy, Elektra has a closed form of variability, when only the values of CVs can change.
Then changes of behavior are limited to the code already present in the program.
Elektra has its focus on contextual (not context) features.
Elektra uses an one-branch context tree:
Without placeholders CVs are automatically non-context-aware configuration items.
Additionally, Elektra supports programmer-declared dependencies.

Umuhoza et al.~\cite{umuhoza2015automatic} compared different ways for code generations on mobile development.
Their methods do not have specific support for context.

Williams et al.~\cite{williams2014contextion} and Biegel et al.~\cite{biegel2004pervasive} presented frameworks for development of context-aware applications.
ContextPhone~\cite{raento2005contextphone} is a further prototyping platform for context-aware mobile applications.
Their approaches focus on developing context sensors, which complements our approach.

\section{Conclusion}
\label{conclusion}

In this paper we presented the idea to use persistent contextual values (CVs) as information source whether layers are active.
We discussed several benefits and limitations of the approach:
(1) The activations themselves take context into account.
(2) Persistent CVs can be used to synchronize layers across different applications in different programming languages.

The approach simplifies taking current context into account and sharing context with other applications.
Furthermore it provides support for individual customization.
End-users can add or redefine configuration in specific context.

In the benchmarks, we showed that activation of CVs is not much more expensive than original layer-activation.
Synchronization of all CVs with configuration files, however, is more costly.
Luckily, the overhead is only constant.
In a real-world benchmark we showed that the sync rate barely has an influence in a web-server setup.

\vspace{0.5em}
Our contributions are:

\begin{enumerate}
\item This combination of performance, context awareness and customization is unique to our approach.
\item Elektra enables programmers to use CVs with code generation in multi-threaded
	and multi-process applications. CVs can even be shared across applications.
\item Our implementation is free software and can be downloaded from
	\url{http://www.libelektra.org}.
	It supports mobile development in C++, Java and other languages.
\item In a case study we described our experience with mobile development,
	and analyzed the performance in microbenchmarks and with a web server.
\end{enumerate}

\acks

Many thanks to Franz Puntigam and the anonymous reviewers for a detailed review of this paper.
Additionally, many thanks to all persons contributing to Elektra.

\bibliographystyle{abbrvnat}

\end{document}